# Ultra long range plasmonic waveguides using quasi-two-dimensional metallic layers.

Jonathan Plumridge and Chris Phillips, *Experimental Solid State Group, Physics Department, Imperial College, Prince Consort Road London SW7 2AZ. UK.*

*Abstract*

We calculate the bound plasmonic modes of a "quantum metamaterial" slab, comprised of multiple quasi-two-dimensional electron gas (Q2DEG) layers, whose thickness is much smaller than the optical wavelength. For the first order transverse magnetic (TM) optical and the surface plasmonic modes we find propagation constants which are independent of both the electron density and of the scattering rates in the Q2DEG's. This leads to extremely long propagation distances. In a detailed case study of a structure comprising a slab of GaAs/AlGaAs multiple-quantum-well (MQW) material, we find propagation lengths of 100's of mm. In addition, the electric field enhancement associated with the plasmonic resonance is found to be sufficient to induce the condition of 'strong coupling' between the slab modes and the intersubband transitions in the MQW's.

*1.1. Introduction*

Surface plasmon polaritons (SPPs) are evanescent waves which travel along the boundary between two materials whose dielectric tensors differ in sign[1]. It has long been known that two such SPPs can couple across a thin metal layer, thus producing a single coupled mode whose propagation length is much longer than an individual SPP[2]. These coupled SPPs can be put into the larger framework of bound modes in slab waveguides for dielectric and weakly absorbing thin films[3]. The coupling across the thin metallic layer hybridises the SPP modes into symmetric and or anti-symmetric admixtures, giving modes known as long and short range plasmons respectively (LRP/SRP). For the LRP the mean Poynting vector within the metallic layer is screened out by the large real part of the dielectric tensor of the metal, which leads to long propagation lengths. Even longer LRP propagation lengths can be attained if the real part of the dielectric constant of the metallic layer is exactly equal to zero[4]; in this case the mean Poynting vector inside the metallic layer is zero and the absorption vanishes. The SRP, on the other hand, always has a large fraction of its power within the metallic layer, and is therefore always heavily damped.

Semiconductor structures have a mature growth and fabrication technology. Nano-engineering in semiconductors can lead to quantum confinement and low dimensional structures such as quantum dots, quantum wires, and quantum wells. Moreover, one can accurately control the electron density, and hence the dielectric constant, by doping the semiconductor with donor atoms. The optical response of a quantum well is strongly anisotropic, and can be well described[5] as a quasi-two-dimensional electron gas (Q2DEG): the electrons are free to move as a gas in the plane of the quantum well, but they are restricted to an atomic-like transition (inter-subband transition, ISBT) in the perpendicular direction.

When quantum wells are stacked in layers to form a multiple quantum well structure (MQW), they form a "quantum metamaterial" slab, comprised of Q2DEG material which has a finite optical thickness but has a similar dielectric tensor to the individual Q2DEGs which make it up. So far, the LRP and SRP modes supported by such a slab of Q2DEG material have not, to our knowledge, been investigated.

Furthermore, in a semi-conductor Q2DEG, the electron density can be tuned over a wide range, allowing the dielectric constant to be tuned from both positive to negative values. The result is the ability to generate LRP – type modes (negative dielectric constant) and more conventional slab waveguide modes[6] (positive dielectric constant) which are strikingly similar to each other. The similarity is such that is will be sufficient to focus our discussion on the LRP modes, being careful to note of any minor distinctions when they arise. The electric fields of the LRP mode are perpendicular to the free motion of electrons in the plane of the Q2DEG, so the metallic character of the electron gas of the Q2DEG does not in itself damp the mode. If there is any absorption it will come from the ISBT, but the flexibility of nano-engineering allows this to be minimised by designing the ISBT energy to be very different from the plasmon resonances. In short, the Q2DEG allows the LRP mode to be supported in a way that allows damping to be made arbitrarily close to zero.

The thickness of the Q2DEG can be much less than the wavelength of light, and for the LRP, this has the effect of generating a resonant, compressed mode which strongly enhances the electric field strength, similar to other plasmonic systems[7,8]. If the ISBT energy is now engineered to be degenerate with the LRP mode, the two excitations will couple strongly together to form LRP-ISBT polaritons. This is similar to previous demonstrations of "strong coupling" using planar microcavities[9], and metal surface plasmon polaritions[10], but the Q2DEG system is simpler to realise. This new LRP mode may also be useful in the future as a waveguide for device applications such as Quantum Cascade Lasers (QCLs)[11] and for non-linear optics, for example second harmonic generation and Surface Enhanced Raman Spectroscopy[12].

In the following section we present the calculated propagation characteristics of the various modes supported by a layer with the dielectric properties of a Q2DEG which is much thinner than the characteristic plasmon wavelength. We also consider how these sub-wavelength guided modes can couple to external propagating light fields, and we compute coupling efficiencies for a typical experimental coupling geometry.

In section 3 we present the results from a transfer matrix (TfM) model of an experimentally realisable multiple quantum well (MQW) system. It operates with plasmon modes and ISBT energies corresponding to mid-infrared wavelengths, and we calculate propagation distances and coupling efficiencies using realistic materials parameters for the GaAs/AlGaAs heterostructure system. Finally, we examine the effect of choosing the ISBT to be degenerate with the long range slab waveguide mode, and we analyse the 'strong coupling' regime. Section 4 is a brief conclusion.

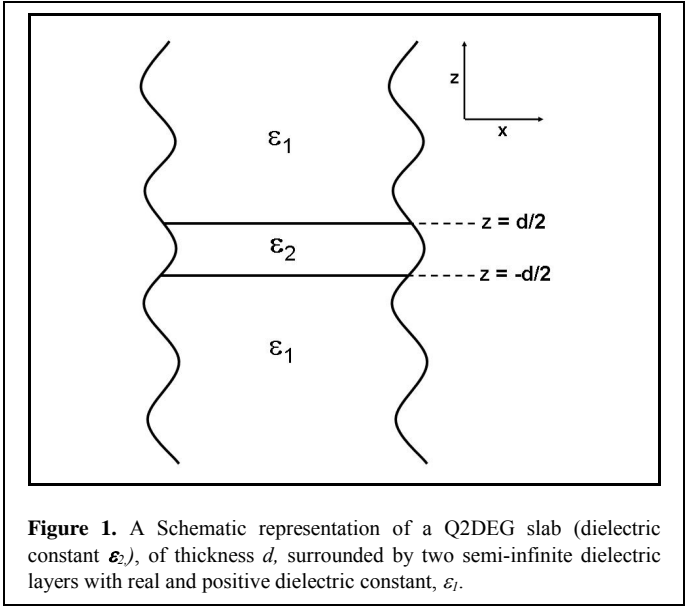

**Figure 1.** A Schematic representation of a Q2DEG slab (dielectric constant $\varepsilon_2$), of thickness $d$, surrounded by two semi-infinite dielectric layers with real and positive dielectric constant, $\varepsilon_1$.

*2.1. Bound plasmon modes supported by a Q2DEG*

We start by considering the general case of a homogenous slab, of thickness $d$ and anisotropic dielectric tensor $\varepsilon_2$, clad by homogenous isotropic dielectrics of dielectric constant $\varepsilon_1$ (figure 1). $\varepsilon_1$ is real and positive whilst $\varepsilon_2$ is written as the tensor:

$$\bar{\varepsilon}_2 = \begin{pmatrix} \varepsilon_{xx} & 0 & 0 \\ 0 & \varepsilon_{yy} & 0 \\ 0 & 0 & \varepsilon_{zz} \end{pmatrix}. \quad (1)$$

Where $\varepsilon_{xx} = \varepsilon_{yy}$ and $\varepsilon_{zz}$ are complex numbers with as yet no restriction. We follow the procedures outlined by Yeh[6] and solve Maxwell's equation for the boundary condition in figure 1 for the modes of linearly polarised light which are bound in the z-direction, but which propagate in the x direction. In the case of the transverse magnetic (TM) polarisation, the modes at the upper and lower slab surfaces couple into two new solutions, the symmetric and the anti-symmetric modes, which are given by equations 2a and 2b respectively.

$$h \tan(hd/2) = q\varepsilon_{xx}/\varepsilon_1; \quad (2a)$$
$$h \cot(hd/2) = -q\varepsilon_{zz}/\varepsilon_1. \quad (2b)$$

Where $h$ and $q$ are defined by:

$$h^2 = K^2 \varepsilon_{xx} - k_x^2 \varepsilon_{xx}/\varepsilon_1, \quad (3)$$
$$q^2 = k_x^2 - K^2 \varepsilon_1, \quad (4)$$

Where $K$ is the free space wavevector, and $k_x$ is the x-component of the wavevector and is known as the complex propagation constant $k_x = k_x^{(r)} - ik_x^{(i)}$. The magnetic fields themselves are given by:

$$H_y(x,z,t) = \exp[i(\omega t - k_x x)] H_m(z) \quad (5)$$

$$H_m(z) = \begin{cases} A\cos(hz) & ; |z| < d/2 \\ B\exp(-qz) & ; z > d/2 \\ B\exp(qz) & ; z < -d/2 \end{cases} \quad (6)$$

The absorption coefficient, $\alpha = 2 k_x^{(i)}$ can now be determined from the imaginary component, $k_x^{(i)}$ of the propagation constant given by equation 5.

There are also, in general, two solutions for linearly polarised light in the x-y plane (the so called tranverse electric or TE modes). Their symmetric and anti-symmetric solutions are:

$$h \tan(hd/2) = q, \quad (7a)$$
$$h \cot(hd/2) = -q, \quad (7b)$$

Respectively, with $h$ and $q$ defined by equations 3 and 4 and with electric fields given by:

$$E_y(x,z,t) = \exp[i(\omega t - k_x x)] E_m(z) \quad (8)$$

$$E_m(z) = \begin{cases} A\cos(hz) & ; |z| < d/2 \\ B\exp(-qz) & ; z > d/2 \\ B\exp(qz) & ; z < -d/2 \end{cases} \quad (9)$$

For both TE and TM polarisations $q$ and $h$ correspond to the z-component of the wavevector $k_z$; $q^2 = -k_z^2$ and $h^2 = k_z^2$. For a bound mode to exist, $q$ must be a positive real number, corresponding to waves that decay in z, for $|z| > d/2$, but there is no such restriction on $h$.

If we now narrow our attention to the case where the slab is composed from one or more semiconductor quantum wells (QWs), each well responds to IR light as a Q2DEG, with a Drude like response to x-y in-plane electric fields, and a Lorentzian oscillator, resonant at the ISBT energy, for the z- component of the electric field. The QW thickness ($L_{QW}$) and the MQW period ($L_{MQW}$) are both much smaller than the plasmon wavelength so we use an effective medium approach to write the components of the slab's dielectric tensor, $\varepsilon_2$, as

$$\varepsilon_{xx} = \varepsilon_{yy} = \varepsilon_y - \left(\frac{\omega_p}{\omega}\right)^2 \frac{1}{1+i\left(\frac{\gamma_1}{\omega}\right)} \quad (10)$$

$$\frac{1}{\varepsilon_{zz}} = \frac{1}{\varepsilon_z} - \left(\frac{\omega_p^2 f_{12}}{2\omega\gamma_2\varepsilon_w}\right) \bigg/ \left[\left(\frac{E_{12}^2 - \hbar^2\omega^2}{2\hbar^2\gamma_2\omega}\right) - i\right]. \quad (11)$$

Where $\varepsilon_y$ and $\varepsilon_z$ are the mean effective background dielectric constants, given by $\varepsilon_y = (1-L_{qw}/L_{mqw})\varepsilon_b + (L_{qw}/L_{mqw})\varepsilon_w$ and $\varepsilon_z^{-1} = (1-L_{qw}/L_{mqw})/\varepsilon_b + (L_{qw}/L_{mqw})/\varepsilon_w$, where $L_{qw}$ and $L_{mqw}$ are the QW width and MQW period respectively, and $\varepsilon_b(\varepsilon_w)$ are the background dielectric constants corresponding to undoped barrier (well) materials. The plasma frequency is $\omega_p = [n_s e^2 / m^* \varepsilon_0 L_{mqw}]^{1/2}$, where $n_s$ is the areal electron density per QW, $m^*$ the electron effective mass and the other symbols have their usual meanings.

The ISBT "artificial atom" Lorentzian oscillator (eq. 11), has energy $E_{12}$, and an oscillator strength $f_{12} = 2m^* E_{12} |z_{12}|^2/\hbar^2$. Here $z_{12} = <1|z|2>$ is the intersubband dipole matrix element and $\gamma_2$ the electron scattering rate, assumed here to be the same in $z$ as in $x$ and $y$. Although we assume that the QW widths and separations are much less than the optical wavelength $[L_{mqw}, L_{qw} << 2\pi c (\varepsilon_{zz})^{-1/2}]$, the total thickness, $d= N*L_{mqw}$ of the slab containing N wells, need not be.

*2.1.1. Ultra long propagation of symmetric LRP modes in the thin film, weak absorber limit.*

We now consider the specific case where the slab of MQW material is much thinner than the optical wavelength, so the 'thin film' limit *(hd/2<<1)* holds.

For the moment we assume that the QW thickness has been chosen to give an ISBT energy very different from that of the plasmon so that absorption of z-polarised fields in the slab is small ($\varepsilon_{zz}^{(i)} << 1$, the "weak absorber" limit). Under these assumptions, for TM polarised modes, the real and imaginary components of the propagation constant, $k_x$, for the symmetric mode (equation 2a) reduce to (see appendix A):

$$k_x^{(r)} = K\sqrt{\varepsilon_1}\left\{1 + \frac{\varepsilon_1}{2}\left[\frac{\pi d}{\lambda}\right]^2 \frac{\left(\varepsilon_{zz}^{(r)} - \varepsilon_1\right)^2}{\left(\varepsilon_{zz}^{(r)}\right)^2}\right\} \quad (12)$$

$$k_x^{(i)} = K\sqrt{\varepsilon_1}\left[\frac{\pi d}{\lambda}\right]^2 \frac{\varepsilon_1^2 \left(\varepsilon_{zz}^{(r)} - \varepsilon_1\right)\varepsilon_{zz}^{(i)}}{\left(\varepsilon_{zz}^{(r)}\right)^3}. \quad (13)$$

We see that now the propagation constant of the symmetric mode is independent of $\varepsilon_{xx}$, viz. the part of the $\varepsilon_2$ tensor that is due to the metallic in-plane electron motion. However, $\varepsilon_{xx}$, and hence the 2D electron density, does determine the *shape* of the mode, (see equations 3-6) and in this thin film, weak absorber limit there are two possible TM symmetric mode profiles that are supported, namely: (i) an LRP-type mode, whose field intensity decays with distance into the slab, akin to that seen in metal films[2-4] and which exists when the real part of $\varepsilon_{xx}$ is negative and (ii) the more conventional slab waveguide mode (denoted TM$_0$, see reference 6), akin to that seen in a dielectric slab waveguide, which exists when the real part of $\varepsilon_{xx}$ is positive (see table 1 for all the modes supported by a Q2DEG in the thin film limit).

| **Modes** | $\varepsilon_{xx}^{(r)}$ | $\varepsilon_{zz}^{(r)}$ | **Polarisation** |
|---|---|---|---|
| LRP | $\varepsilon_{xx}^{(r)} < 0$ | $\varepsilon_{xx}^{(r)} > \varepsilon_1$ | TM |
| SRP | $\varepsilon_{xx}^{(r)} < 0$ | none | TM |
| TM$_0$ | $\varepsilon_{xx}^{(r)} > 0$ | $\varepsilon_{xx}^{(r)} > \varepsilon_1$ | TM |
| TE$_0$ | $\varepsilon_{xx}^{(r)} > \varepsilon_1$ | none | TE |

Table 1. The four types of bound modes (LRP, SRP, TM$_0$ and TE$_0$), supported by a slab of Q2DEG which is symmetrically surrounded by a material whose dielectric constant is both real and positive. The mode type depends upon the value of the dielectric tensor for the Q2DEG ($\varepsilon_2$) and the surrounding material ($\varepsilon_1$).

The beneficial effects of the anisotropy of the Q2DEG can now be clearly seen. The profiles of the LRP/TM$_0$ modes are determined by the 2D electron density, whereas the propagation constant (and hence absorption coefficient) is determined by the weak, z-polarised, absorption of the ISBT. This means that choosing the MQW growth parameters carefully can produce nano-engineered semiconductor structures which will support either LRP, or TM$_0$, type modes, which propagate with low, or negligible, loss.

*2.1.2. The symmetric TE polarised slab waveguide modes in the thin film, weak absorber limit*

The electric field of the TE polarised light is in the x-y plane of the Q2DEG so, as well as taking the thin film approximation, *(hd/2<<1)*, we here also assume that the scattering rates of the electrons in the two dimensional electron gas are low enough to satisfy the $\varepsilon_{zz}^{(i)} << 1$, "weak absorber" condition. This would correspond to experiments with e.g. high mobility modulation doped Q2DEG layers at low temperatures. We then follow the same procedure outlined in 2.1.1. to solve equation (7a), and obtain the following components for the complex

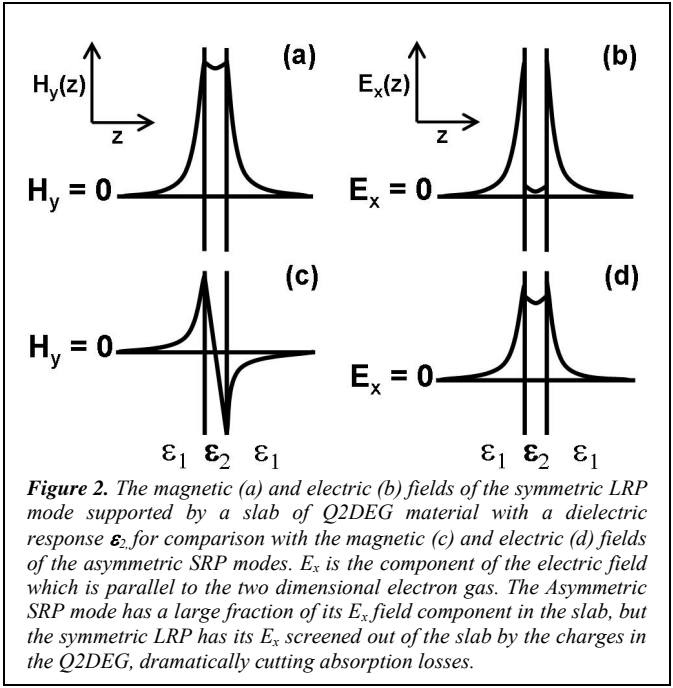

*Figure 2. The magnetic (a) and electric (b) fields of the symmetric LRP mode supported by a slab of Q2DEG material with a dielectric response $\varepsilon_2$, for comparison with the magnetic (c) and electric (d) fields of the asymmetric SRP modes. $E_x$ is the component of the electric field which is parallel to the two dimensional electron gas. The Asymmetric SRP mode has a large fraction of its $E_x$ field component in the slab, but the symmetric LRP has its $E_x$ screened out of the slab by the charges in the Q2DEG, dramatically cutting absorption losses.*

propagation constant of the TE polarised slab waveguide mode (denoted $TE_0$[6]):

$$k_x^{(r)} = K\sqrt{\varepsilon_1}\left\{1 + \frac{1}{2\varepsilon_1}\left[\frac{\pi d}{\lambda}\right]^2 \left(\varepsilon_{xx}^{(r)} - \varepsilon_1\right)^2\right\}, \quad (14)$$

$$k_x^{(i)} = K\sqrt{\varepsilon_1}\frac{\varepsilon_{zz}^{(i)}}{\varepsilon_1}\left[\frac{\pi d}{\lambda}\right]^2 \left(\varepsilon_{xx}^{(r)} - \varepsilon_1\right). \quad (15)$$

The criteria that the various dielectric components need to satisfy in order to give a mode that is bound to the slab are given in table 1. Note that, in this TE polarisation, isolated dielectric interfaces do not support SPP-like modes. This has the consequence that there are no LRP-type modes, (i.e. those whose fields decrease with distance into the slab), in this TE polarisation, because these originate from admixtures of the SPP's on opposing slab faces[6].

For the $TE_0$ mode, both the mode profile and complex propagation constant are determined by the doping density through $\varepsilon_{xx}$. The polarisation of these modes means that they couple only to the in-plane electron response, so they do not "see" the anisotropy designed into the MQW slab. Their mode profiles and complex propagation constants are the same as for a bulk metal slab with the same electron density and mobility and they are therefore always strongly damped.

*2.1.3. The antisymmetric SRP mode in the thin film, weak absorber limit*

Starting with the solution to anti-symmetric equation 2b, for TM polarisation in the thin film *(hd/2<<1)*, weak absorber $\varepsilon_{zz}^{(i)} << 1$ limit, we follow the same arguments as in 2.1.1 , giving the complex propagation constant:

$$k_x^{(r)} = K\sqrt{\varepsilon_1}\left\{1 + \frac{\varepsilon_1}{1\left(\varepsilon_{xx}^{(r)}\right)^2}\left[\frac{\pi d}{\lambda}\right]^{-2}\right\}, \quad (16)$$

$$k_x^{(i)} = -K\sqrt{\varepsilon_1}\left[\frac{\pi d}{\lambda}\right]^{-2}\frac{\varepsilon_1 \varepsilon_{xx}^{(i)}}{\left(\varepsilon_{xx}^{(r)}\right)^3}. \quad (17)$$

Both the complex propagation constant, and mode profile are related to $\varepsilon_{xx}$, the metallic component of the Q2DEG response, so again, these modes do not experience the anisotropy of the Q2DEG. As was the case with the symmetric TE modes, these antisymmetric TM SRP modes have similar absorption and propagation properties to the SRP modes in an equivalent 3D metallic layer, and barely propagate.

The degree to which the various modes will propagate can also be understood by examining their various field distributions. In the "weak-absorber" limit, the ISBT energy has been designed to be very different from the energy of the plasmon mode, so the Q2DEG slab can only absorb energy only through electron motion in the x-y plane. This means that the damping rates will be determined by the x-components of the modes' electric fields inside the slab. These are given by

$$E_x(x,z,t) = -\frac{i}{\omega\varepsilon_0\varepsilon_{zz}}\frac{\partial}{\partial z}H_y(x,z,t). \quad (18)$$

In fig. 2, we clearly see the antisymmetric SRP mode has $E_x$ peaking inside the slab [fig 2(d)], leading to heavy absorption. In the case of the symmetric LRP solution however, the symmetric distribution of charge at the slab surfaces has screened $E_x$ out of the slab [fig. 2 (b)], dramatically decreasing the losses.

*2.1.4. Thick films and higher order modes*

When the layer thickness is increased, so that the "thin film" approximation breaks down, higher order TE and TM modes start to be supported. These have nodes inside the slab, and hence electric fields in the slab which have a larger average component in the x-direction (see equation 18). This has effect of increasing the interaction of the mode with the $\varepsilon_{xx}$ part of the Q2DEG response, and the higher the mode order the larger this effect. By the time the slab thickness approaches the optical wavelength, *(hd/2 ~1)*, the damping reduction benefits, arising from the optical anisotropy of the Q2DEG are all but annulled, and the slab behaves much as would a bulk metal slab with the same mean electron concentration and mobility.

*2.3. Optical coupling schemes for efficiently accessing the propagating LRP mode.*

The plasmonic nature of this propagating LRP mode results in a very compact field structure which is difficult to couple into from a free-space propagating beam using conventional mode-matching techniques. Focussing light into a diffraction-limited spot on one end of the slab waveguide for example, would give a very low coupling efficiency. Here we consider the alternative of using the material surrounding the Q2DEG as an analogue of the prism coupling scheme commonly used in semiconductor waveguide experiments[13] and we analyse how power is transferred into the bound LRP mode which is guided by the slab.

Figure 3 is a schematic of such a "prism coupling" scheme that could easily be manufactured using conventional semiconductor technology, where the substrate dielectric constant is larger than the epilayers' ($\varepsilon_3 > \varepsilon_1$), and $\theta_0$ is greater than the critical angle. The mean power flow from the incident optical beam into the Q2DEG along the z-direction is given by the z-component of the mean Poynting vector:

$$S_z = \begin{cases} 0 \\ -1/2 \operatorname{Re}\left[\dfrac{ih|B|^2 \sin(hz)\cos(hz)}{\omega \varepsilon_0 \varepsilon_{xx}}\right] \\ 0 \end{cases} \begin{array}{l} ; z > d/2 \\ ; |z| < d/2 \\ ; z < -d/2 \end{array} \quad (19)$$

From which we can determine the power flow along the mode itself, i.e. the Poynting vector $S_x$

$$S_x = \dfrac{k_x}{2\omega\varepsilon_0} \begin{cases} \operatorname{Re}\left[|B|^2 \cos(hz)/\varepsilon_{zz}\right] \\ \operatorname{Re}\left[|B|^2 \cos^2(hd/2)\exp(qd - 2qz)/\varepsilon_1\right] \\ \operatorname{Re}\left[|B|^2 \cos^2(hd/2)\exp(qd + 2qz)/\varepsilon_1\right] \end{cases}$$
$$\begin{array}{l} ; |z| < d/2 \\ ; z > d/2 \\ ; z < -d/2 \end{array} \quad (20)$$

B is dependent upon the intensity of the incident plane wave and has a complex dependence on geometrical factors such as the width, $d_1$, of the layer above the MQW slab, and the angle of incidence, $\theta_0$. Analytic solutions for B are tedious to produce and evaluate, even more so if further layers are added, so at this stage we resort to a standard 'transfer matrix' method[14] (TfM, see next section) to provide numerical solutions.

Before we perform this numerical calculation however, it is useful to look at equation 19 in more detail. Inside the Q2DEG slab ($|z| < d/2$) the power flow is dependent upon $\varepsilon_{xx}$ (that part of the dielectric tensor described by the two dimensional electron gas). If $\varepsilon_{xx}$ is a real number, of either sign, then $S_z$ inside the Q2DEG slab will be zero, implying a zero net power flow into the Q2DEG slab; i.e power flowing into the mode will also be able to flow out of it. There will only be a net flow of power into the Q2DEG slab if $\varepsilon_{xx}$ is complex. The z-component of the incident optical power flow corresponds to the energy dissipated by in-plane motion of the electrons in the Q2DEG layers. The x-component of the incident power ends up trapped in the LRP (or $TM_0$) mode, where it subsequently propagates along the slab with the respective absorption coefficient (eq. 13).

*2.4. Numerical Transfer Matrix model applied to a MQW system*

The semiconductor materials that we propose to use (GaAs and AlGaAs) have a mature growth and fabrication technology and well known optical and electronic properties which can be accurately reproduced from one growth run to another[15]. We consider MQW structures which behave as ideal Q2DEGs in the sense that the electrons are perfectly localised in the z-direction by their confining potentials, with no quantum mechanical coupling between neighbouring quantum wells, although of course electrons in different wells still interact via the Coulomb interaction. We also assume that the spacing of the wells is small enough *[$L_{mqw}$, $L_{qw}$ << $2\pi c\, (\varepsilon_{zz})^{-1/2}$ ]*, to allow the use of the 'effective medium' approach and describe the slabs optical response with the $\boldsymbol{\varepsilon}_2$ tensor of equations 10 and 11

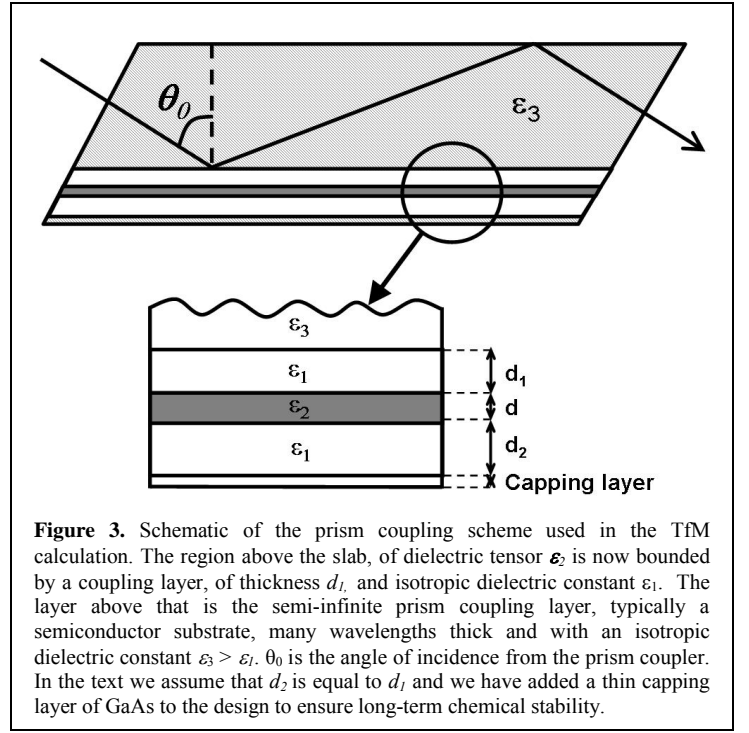

**Figure 3.** Schematic of the prism coupling scheme used in the TfM calculation. The region above the slab, of dielectric tensor $\boldsymbol{\varepsilon}_2$ is now bounded by a coupling layer, of thickness $d_1$, and isotropic dielectric constant $\varepsilon_1$. The layer above that is the semi-infinite prism coupling layer, typically a semiconductor substrate, many wavelengths thick and with an isotropic dielectric constant $\varepsilon_3 > \varepsilon_1$. $\theta_0$ is the angle of incidence from the prism coupler. In the text we assume that $d_2$ is equal to $d_1$ and we have added a thin capping layer of GaAs to the design to ensure long-term chemical stability.

## 3. Calculated coupling efficiencies, propagation lengths, and strong coupling of LRP and $TM_0$ modes supported by a thin film MQW

Table 1 shows a range of possible modes supported by a Q2DEG, all of which can be realised with a slab of MQW and accessed using the "prism-coupled" scheme of fig. 3. We choose parameters to give a slab that will support only the LRP, $TM_0$, and $TE_0$ modes. Our calculations are for TM polarised light, and so we do not consider the $TE_0$

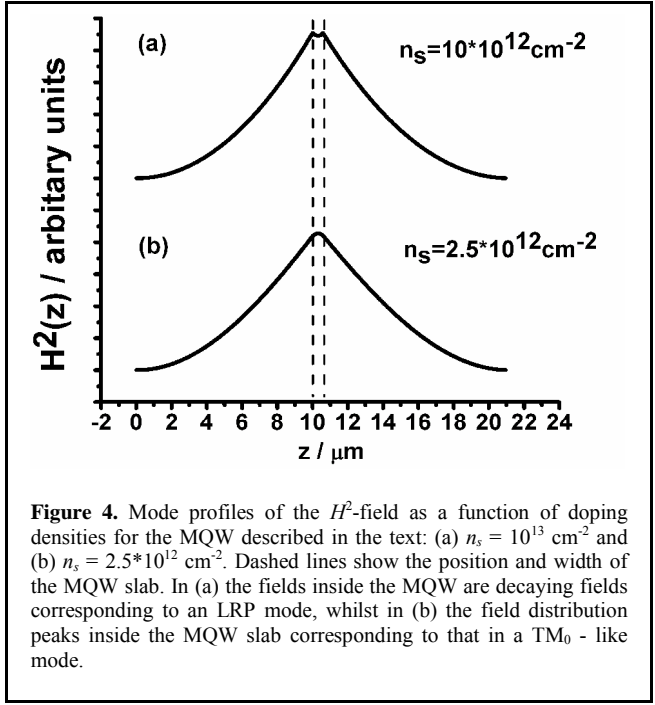

**Figure 4.** Mode profiles of the $H^2$-field as a function of doping densities for the MQW described in the text: (a) $n_s = 10^{13}$ cm$^{-2}$ and (b) $n_s = 2.5*10^{12}$ cm$^{-2}$. Dashed lines show the position and width of the MQW slab. In (a) the fields inside the MQW are decaying fields corresponding to an LRP mode, whilst in (b) the field distribution peaks inside the MQW slab corresponding to that in a $TM_0$ - like mode.

mode.

For the MQW we take 30 repeats ($N$=30) of a quantum well and barrier period, and we use $\varepsilon_b$=9.88, the dielectric constant of Al$_{0.35}$Ga$_{0.65}$As[16], $\varepsilon_3=\varepsilon_w$=10.36 the dielectric constant of GaAs[15] and $\varepsilon_l$=8.2 Al$_{0.9}$Ga$_{0.1}$As[16]. We set $L_{qw}$=6nm and $L_{mqw}$=20nm, $m^*$=0.0665$m_e$. From a separate quantum mechanical calculation of the electron wavefunctions in the QW, we set $z_{21}$=1.4 nm (corresponding to $f_{21}$=0.5), and $E_{21}$=3.75*10$^{14}$s$^{-1}$, which corresponds to a free-space radiation wavelength of $\lambda$= 5 µm. We set $\gamma_1=\gamma_2$=7.596*10$^{12}$s$^{-1}$, which corresponds to the ~ 5 meV ISBT linewidths typically measured[13] in good QW's. These parameter choices set d = 600nm, and we set $d_1$ (unless otherwise stated) = $d_2$ = 10 µm, and we have included a 0.35 µm GaAs capping layer (isotropic dielectric constant $\varepsilon_3$). We now vary only the areal doping density per QW, $n_s$, to explore its effect on the plasmon modes supported by the system as a whole.

### 3.1. Coupling efficiencies into the LRP and $TM_0$ modes of a MQW in the thin film limit

If we consider a beam, of free-space wavelength $\lambda$= 15 µm, it couples to the slab modes when incident at an angle of $\theta_0$=63°. From equation 10, one finds a critical electron density, $n_s$=6.64*10$^{12}$ cm$^{-2}$ above which the modes it couples to have a LRP type character, and below which they are $TM_0$ like (the same result is found by the TfM model). Table 2 lists the various coupling efficiency values and shows a clear correlation with electron density. The higher the electron density, the more effectively the MQW absorbs power from the z-component of the incoming light, thus leaving more power trapped in the LRP, or $TM_0$ mode. This mode then propagates with its power being only weakly absorbed by the ISBT whose energy, in this instance, is 3 times larger than that of the bound mode.

| $n_s$ (10$^{12}$cm$^{-2}$) | Mode | $\alpha$ (mm$^{-1}$) | Coupling Efficiency |
|---|---|---|---|
| 10 | LRP | 0.044 | 36.6% |
| 7.5 | LRP | 0.030 | 25.0% |
| 5 | $TM_0$ | 0.018 | 22.3% |
| 2.5 | $TM_0$ | 0.008 | 9.8% |

Table 2. Coupling efficiencies, computed with the TfM numerical model, for different doping levels, for a $\lambda$= 15 µm free-space beam incident at the right angle to couple into the bound modes of a thin MQW slab. The "absorption coefficients" relate to the in-plane propagation of the resulting slab modes.

The corresponding mode profiles of a LRP ($n_s$=10*10$^{12}$cm$^{-2}$) and a $TM_0$ ($n_s$=2.5*10$^{12}$cm$^{-2}$) mode are plotted in fig.4. Outside the slab, both modes are characterised by similar evanescent field distributions which decay away from the slab. However, the distinguishing feature between the LRP and $TM_0$ modes is the shape of the fields in the MQW slab itself. The LRP mode has clearly originated from the coupling together of the two SPPs supported by the opposite faces of the slab. This is evidenced by the way its $H^2(z)$ field distribution is decaying with distance into the slab.

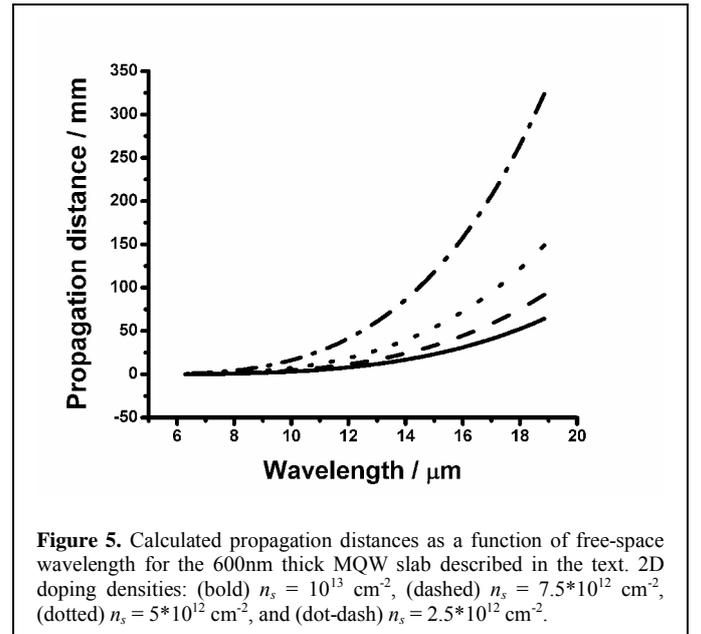

**Figure 5.** Calculated propagation distances as a function of free-space wavelength for the 600nm thick MQW slab described in the text. 2D doping densities: (bold) $n_s = 10^{13}$ cm$^{-2}$, (dashed) $n_s = 7.5*10^{12}$ cm$^{-2}$, (dotted) $n_s = 5*10^{12}$ cm$^{-2}$, and (dot-dash) $n_s = 2.5*10^{12}$ cm$^{-2}$.

*3.2. Propagation lengths of the LRP and $TM_0$ modes of a thin MQW slab calculated with the TfM model.*

For wavelengths between ~6 and 20μm the "thin-film" and "weak absorber" limits are both well obeyed by our 0.6μm thick MQW slab example, for all the modes listed in table 2. As such we can use the analytical treatment (eq. 13) to estimate the absorption coefficient, $\alpha = 2 k^{(i)}_x$ and hence the $1/e$ propagation lengths of the bound slab modes. Figure 5 shows the results, for 4 different doping densities, as a function of mode energy as characterised by the free-space wavelength, $\lambda$, which would excite them. At $\lambda = 15$ μm, i.e. corresponding to the field plots in fig. 4 the propagation distances for the modes in figure 5 a-d are: (a) 24 mm, (b) 34 mm, (c) 55mm, and (d) 121 mm.

As the doping, $n_s$, is increased, the increasing oscillator strength, $f_{21}$, increases the residual ISBT absorption and reduces the propagation lengths. In future though, more refined designs may use more sophisticated QW shapes to reduce $f_{21}$, by band engineering, to counteract this effect. Tuning the ISBT oscillator to a more remote part of the spectrum, by e.g. using narrower wells in a semiconductor

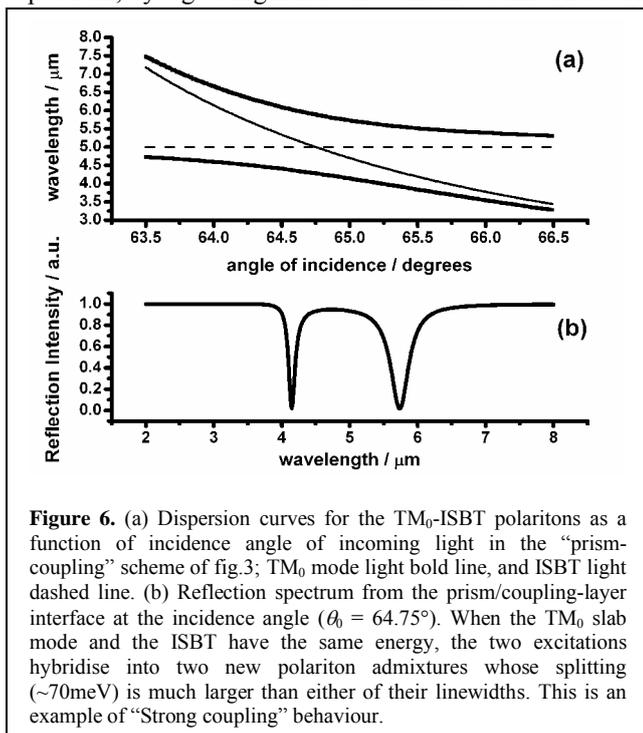

**Figure 6.** (a) Dispersion curves for the $TM_0$-ISBT polaritons as a function of incidence angle of incoming light in the "prism-coupling" scheme of fig.3; $TM_0$ mode light bold line, and ISBT light dashed line. (b) Reflection spectrum from the prism/coupling-layer interface at the incidence angle ($\theta_0 = 64.75°$). When the $TM_0$ slab mode and the ISBT have the same energy, the two excitations hybridise into two new polariton admixtures whose splitting (~70meV) is much larger than either of their linewidths. This is an example of "Strong coupling" behaviour.

system with a larger band offset would have similar advantages.

*3.2. Strong Coupling of an ISBT and $TM_0$ mode*

We have made no assumptions about the values used for the dielectric constants in the TfM model, so we can use it to investigate the point where the ISBT and the $TM_0$ slab mode become degenerate. As there is a strong plasmonic field enhancement around the MQW layer, one would expect an enhancement of non-linear effects such as second harmonic generation, and in the extreme case, the appearance of 'strong coupling'[17]. Strong coupling has been realised in cavity QED experiments[18] with single atoms, and it is characterised by a Vacuum Rabi frequency, $\Omega_{12} = ez_{12}E/\hbar$, (where $E$ is the electric field amplitude associated with the mode and the other symbols have their usual meanings) which is greater than both the linewidth of the atomic transition and the cavity finesse.

In our case the corresponding atomic transition is the ISBT, and the cavity mode is the guided $TM_0$ mode. Plane-wave light incident on the slab couples to different parts of the dispersion curve of the $TM_0$ slab mode according to its angle of incidence. This means that the strength of the coupling between the ISBT and the plasmon mode can be studied by using the TfM model to compute reflectivity spectra, for the experimental scheme of fig. 3, over a range of incidence angles which correspond to tuning the plasmon mode energy through degeneracy with the ISBT.

The result of doing this is very clear anticrossing behaviour (fig.6). At the degeneracy point ($\theta_0 = 64.75°$) two mixed ISBT-$TM_0$ polariton modes appear, which are split by an energy $\hbar\Omega_{12} \sim 70$ meV, i.e. is much larger than their linewidths. For this example the MQW design parameters are as in section 3, except for $n_s = 1.5*10^{12}$ cm$^{-2}$ and $d_1$, which has been reduced to 2.5 μm in the model to increase the coupling efficiency to external beams. This particular design could easily be manufactured by MBE growth techniques.

*4.1. Conclusion*

We have discussed the dispersion relations for the bound plasmon modes supported by a slab whose dielectric response is that of a 2-dimensional electron gas. We find that it supports resonant, compact, low-loss LRP/$TM_0$ – like modes. These can be readily coupled into with free-space optical beams and they can be used to guide optical signals through a patterned semiconductor structure. The ability to separate the in-plane conductivity, needed to support these modes, from the out-of-plane absorption characteristics allows for ultra-long propagation distances to be engineered. A numerical model, seeded with realistic values for the GaAs/AlGaAs semiconductor system, finds propagation distances of the order of 10's cm. The design of the MQW slab and the scheme for optically coupling to its modes are well within the present manufacturing capabilities of MBE growth techniques. Finally, in the simulation, we predict that the degree of electric field enhancement around the MQW will be sufficient to induce 'strong coupling' between a $TM_0$ mode and the ISBT.

*Acknowledgements*

The authors would like to thank both Paul Stavrinou and Robert Steed for stimulating discussions. Funding from the UK Engineering and Physical Sciences Research Council is gratefully acknowledged.

*Appendix A*

The equation for symmetric modes is

$$h \tan(hd/2) = q\varepsilon_{xx}/\varepsilon_1. \qquad (A1)$$

If $d$ is small enough so that $hd/2 << 1$, then this reduces to

$$h^2 d \varepsilon_1 = 2q\varepsilon_{xx}, \qquad (A2)$$

In this thin film limit $d$ is small compared with the optical wavelength, so most of the mode travels in the cladding layers, which makes $k_x^{(r)} \to K\sqrt{\varepsilon_1}$. If we assume the "weak absorber" limit, i.e., that $k_x^{(i)} << 1$, we can define the small quantity $\delta = k_x^{(r)} - K\sqrt{\varepsilon_1}$ which, when substituted into equations (3) and (4) in the main text, gives

$$q = \left[2K\sqrt{\varepsilon_1}\left(\delta + ik_x^{(i)}\right)\right]^{1/2} \qquad (A3)$$

and

$$h = \left[K^2\varepsilon_{xx}\left(1 - \frac{\varepsilon_1}{\varepsilon_{zz}}\right) - q^2\frac{\varepsilon_{xx}}{\varepsilon_{zz}}\right]^{1/2}. \qquad (A4)$$

Now, substituting A4 into A2 gives,

$$\begin{aligned}q^2 &= \frac{1}{4}K^4 d^2 \left(\frac{\varepsilon_1}{\varepsilon_{zz}^{(r)}}\right)^2 \left(\varepsilon_{zz}^{(r)} - \varepsilon_1\right)^2 \\ &+ \frac{i}{2}K^4 d^2 \left(\frac{\varepsilon_1}{\varepsilon_{zz}^{(r)}}\right)^3 \varepsilon_{zz}^{(i)}\left(\varepsilon_{zz}^{(r)} - \varepsilon_1\right)\end{aligned} \qquad (A5)$$

Where we have neglected terms of order $\left(\varepsilon_{zz}^{(i)}\right)^2$ and we have used the condition $\left(\varepsilon_{zz}^{(r)} - \varepsilon_1\right) >> \varepsilon_{zz}^{(i)}$. Equating A5 with A3 now gives the following dispersion relations:-

$$k_x^{(r)} = K\sqrt{\varepsilon_1}\left\{1 + \frac{\varepsilon_1}{2}\left[\frac{\pi d}{\lambda}\right]^2 \frac{\left(\varepsilon_{zz}^{(r)} - \varepsilon_1\right)^2}{\left(\varepsilon_{zz}^{(r)}\right)^2}\right\} \qquad (A6a)$$

$$k_x^{(i)} = K\sqrt{\varepsilon_1}\left[\frac{\pi d}{\lambda}\right]^2 \frac{\varepsilon_1^2\left(\varepsilon_{zz}^{(r)} - \varepsilon_1\right)\varepsilon_{zz}^{(i)}}{\left(\varepsilon_{zz}^{(r)}\right)^3}. \qquad (A6b)$$

*References*

[1] U. Fano, **J. Opt. Soc. Am. 31,** 213 (1941).

[2] C. Quail, J. G. Rako and H. J. Simon, **Opt. Lett. 8,** 377 (1983).

[3] F. Yang, J. R. Sambles and G. W. Bradberry, **Phys. Rev. B. 44,** 4855 (1991).

[4] F. Yang, J. R. Sambles and G. W. Bradberry, **Phys. Rev. Lett. 64,** 559 (1990).

[5] L. Wendler and E. Kändler, **Phys. Status. Solidi. B. 177,** 9 (1993); L. Wendler and T. Kraft, **Phys. Rev. B. 60,** 16603 (1999).

[6] P. Yeh "Optical Waves in Layered Media", John Wiley and Sons (1988).

[7] W. L. Barnes, A. Dereux and T. W. Ebbesen, **Nature. 424,** 824 (2003).

[8] S. A. Maier and H. A. Atwater, **J. Appl. Phys. 98,** 011101 (2005).

[9] E. Dupont, H. C. Liu, A. J. SpringThorpe, W. Lai and M. Extavour, **Phys. Rev. B. 68,** 245320 (2003); D. Dini, R. Köhler, A. Tredicucci, G. Biasiol and L. Sorba, **Phys. Rev. Lett. 90,** 116401 (2003).

[10] M. Zaluzny, W. Zietkowski and C. Nalewajko, **Phys. Rev. B. 65,** 235409 (2002).

[11] C. Sirtori, C. Gmachl, F. Capasso, J. Faist, D. L. Sivo, A. L. Hutchinson and A. Y. Cho, **Opt. Lett. 23,** 1366 (1998).

[12] J. Homola, S. S. Yee and G. Gauglitz, **Sensors. Actuat. B. 54,** 3 (1999).

[13] J F Dynes, M D Frogley, M Beck, J Faist and C C Phillips, **Phys. Rev Lett. 94, 157403 (2005).**

[14] M. Zaluzny and C. Nalewajko, **Phys. Rev. B. 59,** 13043 (1999).

[15] S. Adachi "Properties of Aluminium Gallium Arsenide", INSPEC, the institute of Electrical Engineers (1993).

[16] C. Palmer, P. N. Stavrinou, M. Whitehead and C. C. Phillips, **Semicond. Sci. and Tech. 17,** 1189 (2002).

[17] M. S. Skolnick, T. A. Fisher and D. M. Whittaker, **Semicond. Sci. and Tech. 13,** 645 (1998).

[18] D. M. Sergio "Cavity Quantum Electrodynamics: the strange theory of light in a box", John Wiley and Sons (2005).